\documentclass[aps,reprint,pra]{revtex4-1}
\usepackage[pdftex]{graphicx}
\usepackage{times}
\usepackage{bm}
\usepackage{epsfig}
\usepackage[usenames]{color}

\begin{document}

\title{Recombination process for hydrogen atom in presence of blackbody radiation}

\author{D. Solovyev$^{1}$, T. Zalialiutdinov$^{1}$, A. Anikin$^{1}$, J. Triaskin$^{1}$ and L. Labzowsky$^{1,2}$}

\affiliation{$^1$ Department of Physics, St.Petersburg State University, St.Petersburg, 198504, Russia
}
\affiliation{$^2$ Petersburg Nuclear Physics Institute, 188300, Gatchina, St. Petersburg, Russia}

\begin{abstract}
The process of recombination for the hydrogen atom in the heat bath creating the blackbody radiation is descibed within the frameworks of quantum electrodynamics. For this purpose the self-energy for unbound electron in the field of the nucleus is considered. The imaginary part of this self-energy is directly connected with the recombination cross-section. The same procedure is applied to the hydrogen atom in the field of blackbody radiation. This leads to the new thermal correction to the process of recombination for the hydrogen atom in the heat bath. This correction takes into account the finite lifetimes of atomic levels and appears to be important for special astrophysical studies.
\end{abstract}

\maketitle

\section{Introduction}

The process of recombination and photo-ionization of atoms were intensively studied from the early stages of development of Quantum Mechanics (QM). A description of these processes for the hydrogen atom one can find in \citep{Bethe} and the same for many-electron atoms in \citep{Berest}, \citep{Sob}. The cross-section of these processes are connected via the principle of the detailed balance which for the hydrogen atom looks like
\begin{eqnarray}
\label{1}
\sigma_{nl}^{\rm rec} = 2(2l+1)\frac{k^2}{p^2}\sigma_{nl}^{\rm ion},
\end{eqnarray}
where $k$ is the momentum of the emitted photon, $p$ is the momentum of the incident electron and $nl$ are the quantum numbers (principal and orbital) for the atomic electron state under consideration. In Eq. (\ref{1}) we use the relativistic units ($\hbar=c=m=1$), $m$ is the mass of the electron. The relativistic of Quantum Electrodynamical (QED) description of the photorecombination and photonionization process is also well known (see for example \citep{Sob2}, \citep{Akhiezer}).

In this paper we will apply the QED description of photorecombination and photoionization to the nonrelativistic hydrogen atom. Then the ionization cross-section for the atomic state $nl$ will be expressed as
\begin{eqnarray}
\label{2}
d\sigma_{nl}^{\rm ion} = 2\pi\left|U_{nl,\varepsilon}\right|^2\delta(\omega-\varepsilon-I_{nl})\frac{d^3p}{(2\pi)^3}.
\end{eqnarray}
Here $\omega$ is the incident photon frequency, $\varepsilon$ is the energy of the electron in the continuus spectrum, $I_{nl}$ is the ionization potential for the atomic state $nl$ and $U_{nl,\varepsilon}$ is the amplitude of the process connected with the $S$-matrix element via the relation
\begin{eqnarray}
\label{3}
S_{if}=-2\pi i\delta(E_i-E_f)U_{if},
\end{eqnarray}
where $i$, $f$ denote the initial and final states of the system and $E_i$, $E_f$ are initial and final energies. In the nonrelativistic limit for the hydrogen atom
\begin{eqnarray}
\label{4}
U_{if} = e\sqrt{\frac{2\pi}{\omega}}\left(\vec{e}\,\hat{\vec{p}}\right)_{nl,\varepsilon},
\end{eqnarray}
where $\vec{e}$ is the incident photon polarization vector, $\hat{\vec{p}}$ is the electron momentum operator, $e$ is the electron charge. The cross-section $\sigma_{nl}^{\rm rec}$ can be obtained from Eq. (\ref{2}) with the use of Eq. (\ref{1}). The total photorecombination cross-section then is
\begin{eqnarray}
\label{5}
\sigma^{\rm rec} = \sum\limits_{nl}\sigma_{nl}^{\rm rec}.
\end{eqnarray}

\section{QED evaluation of the photoionization cross-section}

In this paper we employ a different QED approach to the evaluation of the total photorecombination cross-section. This approach is based on the well-known circumstance that the total width of atomic level $\Gamma_{nl}$ equals to sum of transition rates to all the lower-lying atomic states $W_{nl,\,n'l'}$
\begin{eqnarray}
\label{6}
\Gamma_{nl} = \sum\limits_{n'l'\atop (E_{n'l'}<E_{nl})}W_{nl,\,n'l'}.
\end{eqnarray}

\begin{figure}[hbtp]
\includegraphics[scale=0.25]{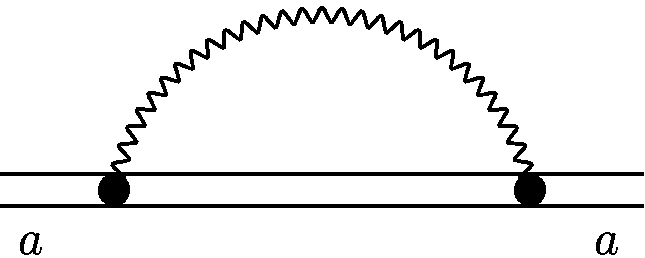}
\caption{The Feynman graph corresponding to the one-loop electron self-energy correction $\Delta E_a$ for the arbitrary atomic state $a$. The double solid line denotes the electron in the field of the nucleus (the Furry picture of QED). The wavy line denotes the virtual photon.}
\label{Fig1}
\end{figure}

In \citep{Klim} it was demonstrated how this principle works within QED description of an atom. The one-loop electron self-energy correction for any atomic state $a$ is presented by the Feynman graph depicted in Fig.~\ref{Fig1}.

According to QED theory \citep{Sob2}, \citep{Akhiezer} the real part of the correction $\Delta E_a$ presented by Fig.~\ref{Fig1} for the bound electron $a\equiv nl$ corresponds to the lowest-order radiative shift (the Lamb shift $L_a$) for the level $nl$. The imaginary part of $\Delta E_a$ represents the total width $\Gamma_{nl}$ of the level $nl$. So
\begin{eqnarray}
\label{7}
\Delta E_{nl}=L_{nl}-\frac{i}{2}\Gamma_{nl}.
\end{eqnarray}
The real part is divergent and should be renormalized. The imaginary part is finite and can be evaluated directly using the Feynman correspondence rules within the Furry picture of QED.

This was done in \cite{Klim}, where the closed relativistic expression for $\Gamma_{nl}$ was obtained and it was demonstrated how the relation (\ref{6}) arises in the nonrelativistic limit. Our idia in the present paper is to obtain in a similar way the closed QED expression for $\sigma^{\rm rec}$ and then to use it for describing the atomic level broadening in the field of blackbody radiation (BBR). For this purpose we define an expression corresponding to Fig~\ref{Fig1} in case when the atomic state $a$ belongs to continous spectrum 
\begin{eqnarray}
\label{8}
\Delta E_{\varepsilon}=L_{\varepsilon}-\frac{i}{2}\Gamma_{\varepsilon}.
\end{eqnarray}
Here $\varepsilon$ denotes the energy of an unbound electron in the field of the nucleus. The radiative shift of the energy $L_\varepsilon$ for the electron in the continous spectrum is out of the scope of the present paper. For the free electron $L_\varepsilon=0$ due to the renormalization. For the free electron the imaginary part of $\Delta E_\varepsilon$ also vanishes, $\Gamma_\varepsilon=0$. This happens because the radiation for the free electron is forbidden by the kinematics. However, for the unbound electron in the field of the nucleus the radiation is not forbidden: the electron can emit photon and transit to the one of the bound states. Consequently
\begin{eqnarray}
\label{9}
\Gamma_\varepsilon=\sigma^{\rm rec}(\varepsilon).
\end{eqnarray}

It is convinient to normalize the incident electron wave function in such a way, that the flux for the incident electron is equal to unity \cite{Sob}. Then the recombination cross-section coincides with the electron capture probability (transition rate) and has the same dimensionality as energy in relativistic units.

According to the Feynman rules in the Furry picture of QED a general nondiagonal second order $S$-matrix element corresponding to Fig.~\ref{Fig1} with $a\equiv \varepsilon$ looks like
\begin{eqnarray}
\label{10}
\langle\varepsilon'|S^{(2)}|\varepsilon\rangle = e^2\int d^4x_1 d^4x_2 \bar{\psi}_{\varepsilon'}(x_1)\gamma^\mu \times
\\
\nonumber
S(x_1,x_2)\gamma^\nu\psi_{\varepsilon}(x_2)D_{\mu\nu}(x_1 x_2),
\end{eqnarray}
where $\psi_{\varepsilon}(x)$ is the Dirac wave function for the unbound electron in the field of the nucleus, $\bar{\psi}$ is the Dirac conjugated wave function, $\gamma^\mu$ are the Dirac matrices, $S(x_1 x_2)$ is the electron propagator for the electron in the Furry picture, $D_{\mu\nu}(x_1 x_2)$ is the photon propagator. We use the standard expressions for $\psi_a(x)$, $S(x_1 x_2)$ and $D_{\mu\nu}(x_1 x_2)$ \cite{Akhiezer}, \cite{Klim}:
\begin{eqnarray}
\label{11}
\psi_\varepsilon(x) = \psi_\varepsilon(\vec{r})e^{-i\varepsilon t},
\end{eqnarray}
the eigenmode decomposition for the electron propagator
\begin{eqnarray}
\label{12}
S(x_1 x_2) = \frac{1}{2\pi i}\int\limits_{-\infty}^\infty d\omega e^{I\omega(t_1-t_2)}\sum\limits_n\frac{\psi_n(\vec{r}_1)\bar{\psi}_n(\vec{r}_2)}{E_n(1-i0)-\omega},\qquad
\end{eqnarray}
where sum runs over the entire dirac spectrum for atomic electron, $E_n$ being the level energies, and the photon propagator in the Feynman gauge
\begin{eqnarray}
\label{13}
D_{\mu\nu}(x_1 x_2) = \frac{1}{2\pi i}\frac{g_{\mu\nu}}{r_{12}}\int\limits_{-\infty}^\infty d\omega e^{i\omega(t_1-t_2)+i|\omega| r_{12}},
\end{eqnarray}
where $r_{12} = |\vec{r}_1-\vec{r}_2|$ and $g_{\mu\nu}$ is the metric tensor.

For the evaluation of the energy shift $\Delta E_\varepsilon$ we use Eq. (\ref{3}) and
\begin{eqnarray}
\label{14}
\Delta E_a = \langle a|U|a\rangle,
\end{eqnarray}
which is valid for irreducable Feynman graphs like Fig.~\ref{Fig1} \cite{Klim}. Inserting Eqs. (\ref{10})-(\ref{13}) into Eqs. (\ref{3}), (\ref{14}), and integrating over the time and frequency variables we find
\begin{eqnarray}
\label{15}
\Delta E_\varepsilon = \frac{e^2}{2\pi i}\sum\limits_n\left(\frac{1-\vec{\alpha}_1\vec{\alpha}_1}{r_{12}}I_{\varepsilon n}(r_{12})\right)_{\varepsilon n n \varepsilon},
\end{eqnarray}
\begin{eqnarray}
\label{16}
I_{\varepsilon n}(r_{12}) = \int\limits_{-\infty}^\infty\frac{e^{i|\omega|r_{12}}d\omega}{E_n(1-i0)-\varepsilon+\omega},
\end{eqnarray}
\begin{eqnarray}
\label{17}
\left(\hat{A}(12)\right)_{abcd}\equiv\langle\bar{\psi}_a(1)\bar{\psi}_b(2)|\hat{A}(12)|\psi_c(1)\psi_d(2)\rangle.
\end{eqnarray}
Here $\vec{\alpha}_i$ are the Dirac matrices acting on the functions depending on the variables $i=1,2$.

To evaluate the $\sigma^{\rm rec}$ according to Eq. (\ref{9}) we need to consider the imaginary part $\Delta E_\varepsilon$, i.e. the real part of the integral $I_{\varepsilon n}(r_{12})$. We rewrite this intgral in the form
\begin{eqnarray}
\label{18}
I_{\varepsilon n}(r_{12}) = \int\limits_{-\infty}^\infty\frac{e^{i\omega r_{12}}d\omega}{E_n-\varepsilon+\omega\mp i0}
\\
\nonumber
-2i\int\limits_{-\infty}^0\frac{\sin\omega r_{12}d\omega}{E_n-\varepsilon+\omega\mp i0}.
\end{eqnarray}
In the second intgral in Eq. (\ref{18}) we can omit $\mp i0$ in the denominator since this denominator has no zeros within the interval $(-\infty,0]$. Then the second term in the right-hand side of Eq. (\ref{18}) is pure imaginary and we will concentrate at first term. The sign $\mp$ in the denominator of the first integral corresponds to $E_n<0$, $E_n>0$. Evaluating this integral in the complex plane we close the contour of intgration in the upper hals-plane since in this half-plane the intgral vanishes along the half-circle with infintely large radius. The existence of the pole inside this contour depends on the sign of $E_n$: the pole exists when $E_n>0$. Evaluation of residue in this pole gives
\begin{eqnarray}
\label{19}
{\rm Re} I_{\varepsilon n}(r_{12})=2\pi \sin\left((\varepsilon-E_n)r_{12}\right)
\end{eqnarray}
and finally we obtain the closed expression for $\Gamma_\varepsilon$:
\begin{eqnarray}
\label{20}
\Gamma_\varepsilon = -2{\rm Im} \Delta E_\varepsilon = \qquad\qquad\qquad
\\
\nonumber
=2 e^2\sum\limits_{n\atop E_n>0}\left(\frac{1-\vec{\alpha}_1\vec{\alpha}_2}{r_{12}}\sin\left((\varepsilon-E_n)r_{12}\right)\right)_{\varepsilon n n\varepsilon}.
\end{eqnarray}
This expression is an analog of the expression for the full radiative one-photon width for the one-electron atom derived in \cite{Klim}:
\begin{eqnarray}
\label{21}
\Gamma_a = 2e^2\sum\limits_{n, E_n<E_a\atop E_n>0}\left(\frac{1-\vec{\alpha}_1\vec{\alpha}_2}{r_{12}}\sin\left((E_a-E_n)r_{12}\right)\right)_{a n n a}.
\end{eqnarray}

The value $\Gamma_\varepsilon$ also represents the one-photon (lowest order in $e$) contribution to the photorecombination for the one-electron atom. In a similar way an expression for the cross-section of bremsstarlung for the electron scattered on the nucleus can be derived
\begin{eqnarray}
\label{22}
\Gamma^{\rm BS}_\varepsilon 
=2 e^2\int\limits_0^\varepsilon d\varepsilon' \left(\frac{1-\vec{\alpha}_1\vec{\alpha}_2}{r_{12}}I_{\varepsilon\varepsilon'}(r_{12})\right)_{\varepsilon \varepsilon' \varepsilon'\varepsilon}.
\end{eqnarray}
Expressions (\ref{20}), (\ref{21}), (\ref{22}) are valid foe one-electron ions with arbitrary high value of the nuclear charge Z. Evaluation of the level widths ot cross-sections for the highly charged ions with use of these formulas is out of scope of the present paper.

\section{QED evaluation for the photorecombination for an atom in a heat bath}

Our goal in the present paper is to evaluate the photorecombination cross-section for the hydrogen atom in the field of BBR by the method described in section II. When describing the processes of photonionization and photorecombination in atoms the influence of the stimulated emission also hass to be taken into account \cite{Sob}. In a sense, the BBR also provides the stimulated emission with the Planck frequency distribution. The influence of the BBR on $\sigma^{\rm rec}$ and $\sigma^{\rm ion}$ is of interest also for astrophysics \citep{pejo}, \citep{Boardman} since the Cosmic Microwave Background (CMB) possesses the properties of BBR. A QED description of the heated electron-photon plasma was made for example in \citep{Don}, \citep{DHR}, where the free electron and photon propagators in the field of BBR were derived. The application of QED theory to the atoms in the field of BBR was recently made in \citep{SLP-QED}. The photon propagators in \citep{SLP-QED} were assumed to be the sam as in \citep{Don}, \citep{DHR}, but the electron propagators were the same as for the bound electrons in atoms. The reason was that the BBR field can essentially destroy the atomic structure only at the very high temperatures. In the present paper we assume that the same remarks are valid for the unbound electrons in the field of the nucleus. The latter field is still considered to be much stronger than the BBR field.

In \citep{SLP-QED} the same one-loop electron self-energy as depicted in Fig.~\ref{Fig1} was applied for the description of the radiative shift for atomic electrons in the field of the BBR. The only difference with the description in section II of this paper was that the photon propagator Eq. (\ref{13}) was replaced by propagator for thermal photons with the Planck frequency distribution. An imaginary part of this radiative shift was considered as the atomic level broadening in the field of the BBR. In this section we will do the same for the unbound electron in the field of the nucleus. This will give us the correction to $\sigma^{\rm rec}$ for atoms in the heat bath.

The photon propagator for the thermal photons derived in \citep{Don}, \citep{DHR} looks like
\begin{eqnarray}
\label{23}
D_{\mu\nu} = -4\pi g_{\mu\nu}\int\frac{d^4k}{(2\pi)^4}n_\beta(|\vec{k}|)e^{ik(x_1-x_2)}\delta(k^2),
\end{eqnarray}
where $k\equiv(\vec{k},\omega)$, $k^2=\vec{k}^2-\omega^2=|\vec{k}|^2-\omega^2$, $n_\beta(|\vec{k}|)=(e^{\beta|\vec{k}|}-1)^{-1}$, $\beta = k_B T$, $k_B$ is the Boltzman constant and $T$ is the radiation temperature. An expression for the thermal radiative shift for the unbound electron in the field of the nucleus is similar to Eq. (\ref{15}):
\begin{eqnarray}
\label{24}
\Delta E_\varepsilon^\beta = \frac{e^2}{\pi}\sum\limits_{n\atop (|E_n|<I)}\left(\frac{1-\vec{\alpha}_1\vec{\alpha}_2}{r_{12}}I_{n\varepsilon}^\beta(r_{12})\right)_{\varepsilon nn \varepsilon},
\end{eqnarray}
where
\begin{eqnarray}
\label{25}
I_{n\varepsilon}^\beta(r_{12}) = 2\int\limits_{-\infty}^\infty d\omega\int\limits_0^\infty d|\vec{k}||\vec{k}|\sin\left(|\vec{k}|r_{12}\right)\frac{\delta(|\vec{k}|^2-\omega^2)n_\beta(|\vec{k}|)}{E_n(1-i0)-\varepsilon+\omega}.\qquad
\end{eqnarray}
The limitation $|E_n|<I$ for the sum over $n$ in Eq. (\ref{24}), where $I$ is the ionization potential for the ground state of an atom, means that the summation is carried out only over the discrete spectrum.

For obtaining Eq. (\ref{25}) the integration over angles in Eq. (\ref{23}) was performed. Presenting $\delta$-function in Eq. (\ref{25}) in the form
\begin{eqnarray}
\label{26}
\delta(|\vec{k}|^2-\omega^2)=\frac{1}{2|\vec{k}|}\left[\delta(\omega+|\vec{k}|)+\delta(\omega-|\vec{k}|)\right]
\end{eqnarray}
and integrating over $\omega$ in Eq. (\ref{25}) we find
\begin{eqnarray}
\label{27}
I_{n\varepsilon}^\beta(r_{12}) = \int\limits_0^\infty d|\vec{k}|\sin\left(|\vec{k}|r_{12}\right)n_\beta(|\vec{k}|)\times
\\
\nonumber
\left[\frac{1}{E_n(1-i0)-\varepsilon+|\vec{k}|}+\frac{1}{E_n(1-i0)-\varepsilon-|\vec{k}|}\right].
\end{eqnarray}

For proceeding further we can use the "pole approximation" in Eq. (\ref{27}). Then only the first term in square brackets in Eq. (\ref{27}) contributes. Formally, this contribution can be obtained via the Sokhotski-Plemelj relation
\begin{eqnarray}
\label{28}
\frac{1}{x-i\epsilon} = P.V.\frac{1}{x}+i\pi\delta(x),
\end{eqnarray}
where $P.V.$ means that when integrating the left part of Eq. (\ref{28}) with any complex (analytical) function the integral with the first term in the right-hand part of Eq. (\ref{28}) should be understood as a principal value integral. Note also that considering the thermal corrections we limit ourselves only with positive values of $E_n$. Since we are interested in imaginary part of $\Delta E_\varepsilon^\beta$ and hance of $I_{n\varepsilon}^\beta(r_{12})$ we obtain the following results
\begin{eqnarray}
\label{29}
{\rm Im} I_{n\varepsilon}^\beta(r_{12}) = \pi \sin\left((\varepsilon-E_n)r_{12}\right)n_\beta(\varepsilon-E_n),
\end{eqnarray}
\begin{eqnarray}
\label{30}
{\rm Im} \Delta E_\varepsilon^\beta = e^2\sum\limits_n\left(\frac{1-\vec{\alpha}_1\vec{\alpha}_2}{r_{12}}\sin\left((\varepsilon-E_n)r_{12}\right)n_\beta(\varepsilon-E_n)\right)_{\varepsilon n n \varepsilon}.\qquad
\end{eqnarray}

In the nonrelativistic limit $(\varepsilon-E_n)r_{12}\sim \alpha$ (in r.u., $\alpha$ is the fine structure constant) and we canexpand $\sin\left((\varepsilon-E_n)r_{12}\right)$ in Taylor series. For the term independent on $\vec{\alpha}$-matrices in Eq. (\ref{30}), the first term of Taylor expansion vanishes due to the orthogonality of the vawe functions $\langle \varepsilon|n\rangle=0$. The next term, with the use of relation $r^2_{12}=r_1^2+r_2^2-2(\vec{r}_1\vec{r}_2)$ reduces to the dipole matrix elements $\langle\varepsilon|\vec{r}|n\rangle\langle n|\vec{r}|\varepsilon\rangle$. For the term depending on $\vec{\alpha}$-matrices in Eq. (\ref{30}) the first term of Taylor expansion works. Remebering that in the nonrelativistic limit $\langle\varepsilon|\vec{\alpha}|n\rangle = \langle\varepsilon|\hat{\vec{p}}|n\rangle$, where $\hat{\vec{p}}$ is the momentum operator and using the known quantum-mechanical relation $\langle\varepsilon|\vec{p}|n\rangle = -i(\varepsilon-E_n)\langle\varepsilon|\vec{r}|b\rangle$ we again reduce the $\vec{\alpha}$- matrix contribution to the product $\langle\varepsilon|\vec{r}|n\rangle\langle n|\vec{r}|\varepsilon\rangle$. Combining both contributions we finally arrive at the quantum-mechanical expression for the photorecombination cross-section for an atom in the BBR field \citep{Sob}:
\begin{eqnarray}
\label{31}
\sigma^{{\rm rec},\beta}_\varepsilon = \frac{4}{3}e^2\sum\limits_{n\atop (|E_n|<I)}(\varepsilon-E_n)n_\beta(\varepsilon-E_n)\left|\langle\varepsilon|\vec{p}|n\rangle\right|^2.\qquad
\end{eqnarray}

\section{QED theory of the photorecombination in the heat bath with atomic level widths taken into account}

In our previous work \citep{SLP-QED} it was shown that the atomic level width within the QED approach for bound-bound transitions (i.e. for atomic level broadening) in the field of the BBR leads to the important change in the results. In \citep{SLP-QED} it was also argued that from the other point of view the influence of BBR can be considered as a mixing of atomic levels by the field of the BBR. Recently, this idea was applied to the astrophysics of cosmological recombination \cite{Zal-19}. It was shown that the influence of the BBR (CMB) shifts the values of cosmological parameters by $\sim 1\%$, i.e. as much as the other important corrections. In this section we will apply the same approach to the photoionization process which are the unbound-bound transitions.

Now we have to return to Eq. (\ref{27}) and to insert the atomic level widths within the frames of QED. For this purpose we have to consider the double loop Feynman graph Fig. 2.
\begin{figure}[hbtp]
\includegraphics[scale=0.25]{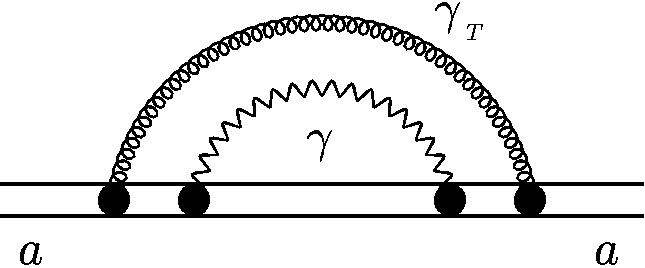}
\caption{The double-loop Feynman graph where the outer photon line corresponds to the thermal photon $\gamma_T$ and the inner photon line corresponds to the ordinary (optical) photon. The other notations are the same as in Fig.~\ref{Fig1}.}
\label{Fig2}
\end{figure}
It is easy to check that after using again the Feynman rules, employing the expression (\ref{12}) for the electron propagator, expression (\ref{13}) for the ordinary photon propagator, expression (\ref{23}) for the thermal photon propagator and after integrating over time and frequency variables in the pole approximation, we will get for $\Delta E_\varepsilon^\beta$ the same expression as (\ref{24}) but with the integral (\ref{27}) replaced by
\begin{eqnarray}
\label{32}
I_{n\varepsilon}^\beta(r_{12}) = \int\limits_0^\infty d\omega_T \sin\left(\omega_t r_{12}\right) n_\beta(\omega_T)\times
\\
\nonumber
\frac{1}{E_n(1-i0)-\varepsilon+\omega_T}\Delta E_n\frac{1}{E_n(1-i0)-\varepsilon+\omega_T}.
\end{eqnarray}

In this section we use the notation $\omega_T=|\vec{k}|$ for the thermal photon frequency. The expression for $\Delta E_n$ is the standard expression for the radiative shift of the atomic level $n$ in the ordinary QED. It can be obtained from Eq. (\ref{15}) by replacing $\varepsilon\rightarrow n$ (the summation index in (\ref{15}) then also has to be changed, $n\rightarrow n'$). The  pole approximation means that only the pole value $\omega_T=\varepsilon-E_n$ contributes to the integral over $\omega_T$ in Eq. (\ref{32}). In this approximation we can neglect the contribution of the other double-loop graphs (crossed loops, loop after loop). Also in the third (from the left to the right) electron propagator in Fig.~\ref{Fig2} we can retain only the term with $n'=n$. In the energy denominators in Eq. (\ref{32}) in the nonrelativistic approximation for the hydrogen atom only the positive values $E_n$ should be retained and in what follows the integration over $\omega_T$ will go along the real axis. 

In the pole approximation we can continue to insert new electron self-energy parts into the internal electron line in Fig.~\ref{Fig2}, every time keeping the same pole denominator in the new electron propagators. All these insertions create a geometric progression. Summation of this progression gives
\begin{eqnarray}
\label{33}
I_{n\varepsilon}^\beta(r_{12}) = \int\limits_0^\infty d\omega_T \frac{\sin\left(\omega_t r_{12}\right) n_\beta(\omega_T)}{E_n(1-i0)-\varepsilon+\omega_T+\Delta E_n},
\end{eqnarray}
where $\Delta E_n = L_n-\frac{i}{2}\Gamma_n$. Further we will neglect the Lamb shift $L_n$ and keep only the level width $\Gamma_n$ in the our calculations. In this way the spectral line profile was first derived within QED \citep{Low}. In \citep{SLP-QED} this way of inserting the atomic level widths was extended outside the pole approximation, for the second term in square brackets in Eq. (\ref{27}) and for all the values of $\omega_T$ in the integral over $\omega_T$ in this expression. This contribution in \citep{SLP-QED}, \cite{Zal-19} was called "nonresonant" unlike the pure pole ("resonant") contribution. With this definition the "nonresonant" approximation includes also the "resonant" part. In \citep{SLP-QED} a procedure, that was originally applied for the resonant photon scattering on an atom, was used for introducing atomic level widths in the denominators in Eq. (\ref{27}). This is strictly justified within QED for the first term in square brackets in Eq. (\ref{27}). In the second term in square brackets the imaginary part in denominator can be omitted since there is no pole. Therefore, the second term in Eq. (\ref{27}) is pure real and does not contribute to the imaginary part of the correction Eq. (\ref{24}). %Anyhow it corresponds to the standard quantum mechanical way of introducing the atomic level widths: every level energy is replaced by $E_n-\frac{i}{2}\Gamma_n$. The introducing of the atomic level width was found recently cricial for the other "nonresonant" processes of interaction of atoms with radiation field \cite{jentlach1}, \cite{jentlach2}.

Our final expression for the photorecombination cross-section looks like
\begin{eqnarray}
\label{34}
\tilde{\sigma}^{{\rm rec}, \beta}_\varepsilon = - 2{\rm Im} \Delta E_\varepsilon^\beta 
=\qquad
\nonumber
\\
\frac{2e^2}{3\pi}\sum\limits_n \left|\langle\varepsilon|\hat{\vec{p}}|n\rangle\right|^2 \int\limits_0^\infty d\omega_T
\frac{ n_\beta(\omega_T)\Gamma_n}{(E_n-\varepsilon+\omega_T)^2+\frac{1}{4}\Gamma_n^2}
%+\frac{\Gamma_n}{(E_n-\varepsilon-\omega_T)^2+\frac{1}{4}\Gamma_n^2}\right]
.\qquad
\end{eqnarray}
We remind that this expression is written in special units when the incident electron current is set equal to unity. Otherwise Eq. (\ref{34}) should be divided by this current. The "nonresonant" expression (\ref{34}) reduces to the "resonant" one Eq. (\ref{31}) when $\Gamma_n=0$. This can be seen with the use of the formula
\begin{eqnarray}
\label{35}
\delta(x)=\frac{1}{\pi}\lim\limits_{\epsilon\rightarrow 0}\frac{\epsilon}{x^2+\epsilon^2}.
\end{eqnarray}

\section{Recombination and ionization coefficients}
The effective cross-sections evaluated in previous sections allows for the definition of recombination and ionization coefficients \cite{Sob2}. The rate of recombination to the $n$-th level due to the spontaneous recombination processes, $\alpha_{nl}$, is given by
\begin{eqnarray}
\label{36}
\alpha_{nl}=\int\limits_{0}^{\infty}\sigma_{nl}^{\rm rec}f(v)vdv,
\end{eqnarray}
where $f(v)$ is the Maxwell-Boltzmann distribution function with the velocity of incident electrons $v$ ($v=p$ in our units):
\begin{eqnarray}
\label{37}
f(v) dv = 4\pi \left(\frac{1}{2\pi k_B T}\right)^{3/2} v^2 e^{-\frac{v^2}{2k_B T_{e}}} dv\,.
\end{eqnarray}
The presence of the Maxwell-Boltzmann distribution function in recombination coefficient restricts the magnitude of the incident electron momentum $p$. The typical speed can be estimated as $p^2\sim 2k_B T\ll 1$ upto $T\sim 10^5$ K what justifies the nonrelativistic approximation used to obtain Eq. (\ref{31}).

The similar equation can be written for the stimulated recombination coefficient
\begin{eqnarray}
\label{38}
\alpha_{nl}^\beta=\int\limits_{0}^{\infty}\sigma_{nl}^{\rm rec,\beta}f(v)vdv,
\end{eqnarray}
and the total recombination coefficient is
\begin{eqnarray}
\label{39}
\alpha^{\rm total}\equiv\alpha_A=\sum\limits_{nl}\alpha_{nl},
\end{eqnarray}
where index $A$ corresponds to the so called case A when the coefficient $\alpha^{\rm total}$ includes the direct recombination process to the ground state, while the case B in astrophysical researches excludes this process.

According to Eq. (\ref{34}) the most curious result arises for the ground state of an atom. Since the natural level width of ground state is zero the cross-section of recombination process reduces to Eq. (\ref{31}), i.e. there is the only resonant contribution. 
The spontaneous recombination coefficients defined by Eqs. (\ref{22}), (\ref{36}) and stimulated recombination coefficients making use of Eqs. (\ref{31}), (\ref{34}) and (\ref{38}) are compared numerically in Table~\ref{tab:1}.
\begin{widetext}
\begin{center}
\begin{table}
\caption{The recombination coefficients for spontaneous and stimulated recombination processes for $1s$, $2s$, $3s$, $5s$ and $10s$ states at different temperatures. Coefficients $\alpha_{nl}$ are calculated with the use of Eq. (\ref{22}), while the coefficients $\alpha_{nl}^\beta$ and $\tilde{\alpha}_{nl}^\beta$ are defined by Eqs. (\ref{31}) and (\ref{34}), respectively. All values are given in $m^3/s$.}	\label{tab:1}
\begin{tabular}{c c c c c c c}
\hline
{} & $T=300$ K &  $T=1000$ K &  $T=3000$ K &  $T=5000$ K &  $T=10000$ K &  $T=20000$ K\\
\hline
$\alpha_{1s}$ & $9.4939\times 10^{-19}$ & $ 5.1848\times 10^{-19} $ & $ 2.9688\times 10^{-19} $ & $2.2812\times 10^{-19}$ & $ 1.5819\times 10^{-19} $  & $ 1.0787\times 10^{-19} $\\

$\alpha_{1s}^\beta$ & $ 0.0 $ & $6.9968\times 10^{-88}$ & $2.0781\times 10^{-42}$ & $2.2263\times 10^{-33}$ & $1.1211\times 10^{-26}$ & $2.0858\times 10^{-23}$\\

\hline
$\alpha_{2s}$ & $1.39195\times 10^{-19}$ & $7.6117\times 10^{-20}$ & $4.3716\times 10^{-20}$ & $3.3664\times 10^{-20}$ & $2.34199\times 10^{-20}$ & $1.5998\times 10^{-20}$\\

$\alpha_{2s}^\beta$ & $5.02703\times 10^{-77}$ & $2.7449\times 10^{-37}$ & $4.2385\times 10^{-26}$ & $6.3229\times 10^{-24}$ & $2.3283\times 10^{-22}$ & $1.2711\times 10^{-21}$\\

$\tilde{\alpha}_{2s}^\beta$  & $6.5603\times 10^{-39}$ & $3.509\times 10^{-37}$ & $4.2385\times 10^{-26}$ & $6.3204\times 10^{-24}$ & $2.3283\times 10^{-22}$ & $1.2711\times 10^{-21}$\\

\hline
$\alpha_{3s}$ & $4.68487\times 10^{-20}$  & $2.5616\times 10^{-20}$ & $1.46942\times 10^{-20}$ & $1.12941 \times 10^{-20}$ &  $7.81646\times 10^{-21}$ & $5.28975\times 10^{-21}$\\

$\alpha_{3s}^\beta$ & $9.4105\times 10^{-46}$ & $3.0844\times 10^{-28}$ & $2.13561\times 10^{-23}$ & $1.74178\times 10^{-22}$ & $7.83262\times 10^{-22}$ & $1.62699\times 10^{-21}$\\

$\tilde{\alpha}_{3s}^\beta$ & $1.8846\times 10^{-32}$  & $3.0855\times 10^{-28}$ & $2.13561\times 10^{-23}$ & $1.7418\times 10^{-22}$ & $7.8326\times 10^{-22}$ &  $1.62699\times 10^{-21}$\\

\hline
$\alpha_{5s}$ & $1.22698\times 10^{-20}$ & $6.67498\times 10^{-21}$ & $3.77965\times 10^{-21}$ & $2.87475\times 10^{-21}$ & $1.95188\times 10^{-21}$ & $1.2908\times 10^{-21}$\\

$\alpha_{5s}^\beta$ & $4.4237\times 10^{-30}$ & $6.07121\times 10^{-24}$ & $2.54593\times 10^{-22}$ & $5.16808\times 10^{-22}$ & $8.74335\times 10^{-22}$ & $1.11702\times 10^{-21}$ \\

$\tilde{\alpha}_{5s}^\beta$ & $4.4687\times 10^{-30}$ & $6.07121\times 10^{-24}$ & $2.54593\times 10^{-22}$ & $5.16808\times 10^{-22}$ & $8.74335\times 10^{-22}$ & $1.11702\times 10^{-21}$\\

\hline
$\alpha_{10s}^\beta$ & $2.06681\times 10^{-21}$ & $1.09008\times 10^{-21}$ & $5.87262\times 10^{-22}$ & $4.34211\times 10^{-22}$ & $2.83431\times 10^{-22}$ & $1.80743\times 10^{-22}$\\

$\alpha_{10s}^\beta$ & $5.42689\times 10^{-24}$ & $1.34303\times 10^{-22}$ & $3.19801\times 10^{-22}$ & $3.74313\times 10^{-22}$ & $4.04424\times 10^{-22}$ & $3.92211\times 10^{-22}$\\

$\tilde{\alpha}_{10s}^\beta$ & $5.42689\times 10^{-24}$ & $1.34303\times 10^{-22}$ & $3.19801\times 10^{-22}$ & $3.74313\times 10^{-22}$ & $4.04424\times 10^{-22}$ & $3.9221\times 10^{-22}$\\
 \hline
  \hline
\end{tabular}
\end{table}
\end{center}
\end{widetext}

It is shown that the stimulated recombination coefficient becomes important for the high temperatures and can exceed the spontaneous one for the highly excited states. The numerical values in Table~\ref{tab:1} obtained with Eq. (\ref{31}) are in a perfect agreement with the quantum mechanical results \cite{pejo}, \cite{Boardman}. The values of stimulated recombination coefficient defined by Eq. (\ref{34}) are significantly larger than the results obtained with Eq. (\ref{31}) at low temperatures for low lying excited states and become comparable with them at temperature about $1000$ K for any states.% Such behavior was noted in \cite{jentlach1}, \cite{jentlach2}. 

%On example of $2s$ state in hydrogen atom the stimulated recombination coefficient evaluated with the QM result Eq. (\ref{31}) ($\alpha^\beta_{2s}=5.02699\times 10^{-77}$ $m^3/s$ at the temperature $300$ K) should be compared with the QED result which is equal to $\tilde{\alpha}^\beta_{2s}=6.56\times 10^{-39}$ $m^3/s$. The latter is significantly larger than QM result. This difference can be explained by the "nonresonant" contribution arising in Eq. (\ref{34}). However, the both results are still negligible with respect to the spontaneous coefficient $\alpha_{2s} = 1.39195\times 10^{-19}$ $m^3/s$. The order of magnitude for stimulated process becomes comparable with the spontaneous process for temperatures larger than $20000$ K. In opposite to the $2s$ state, the stimulated recombination coefficients are the same for highly excited states and become larger than the spontaneous one at lower temperatures with increasing of principal quantum number $n$. 

Numerical results for the total recombination (spontaneous and stimulated) and ionization coefficients (summed over all $nl$ states) are collected in Table~\ref{tab:2}.
\begin{widetext}
\begin{center}
\begin{table}
\caption{The total recombination coefficients for spontaneous, stimulated photorecombination and photoionization processes at different temperatures are listed. Coefficients $\alpha_A$ are calculated by the summation of Eqs. (\ref{22}) and (\ref{36}) over $nl$, while the coefficients $\alpha^\beta_A$ and $\tilde{\alpha}^\beta_A$ are defined by Eqs. (\ref{31}), (\ref{34}) and (\ref{38}), respectively. The total photoionization coefficient, $\beta_A$ is obtained with the use of Eqs. (\ref{1}) and (\ref{22}). These values were obtained by the direct summation over $nl$ upto $n=70$, $n=100$, $n=150$ and $n=300$ (the first, second, third and fourth sublines in rows, respectively). In the last two rows the values of recombination coefficient corresponding to the case B are listed, $\alpha_B$ represents our calculations and $\alpha^{S}_B$ were obtained with the use of Eq. (\ref{40}), see \cite{Seager_2000}.  All values are given in $m^3/s$. The three last sublines in the last row represent the relative difference, $(\alpha_B^{S}-\alpha_B)/\alpha_B^{S}$, in percents.}	\label{tab:2}
\begin{tabular}{ c | c  c  c  c  c  c  c}
\hline
{} & $T=300$ K &  $T=700$ K & $T=1000$ K &  $T=3000$ K &  $T=5000$ K &  $T=10000$ K &  $T=20000$ K\\
\hline
$\alpha_A$ & $4.16281\times 10^{-18}$ & $2.46308\times 10^{-18} $ & $1.96327\times 10^{-18}$ & $9.54546\times 10^{-19}$ & $6.74151\times 10^{-19}$  & $4.14493\times 10^{-19}$ & $2.49899\times 10^{-19}$\\

$n=100$ & $4.23453\times 10^{-18}$ & $2.48994\times 10^{-18} $ & $1.98078\times 10^{-18}$ %$0.9\%$
  & $9.58999\times 10^{-19}$ & $6.76464\times 10^{-19}$  & $4.15429\times 10^{-19}$ & $2.50273\times 10^{-19}$\\
  
$n=150$ & $4.28424\times 10^{-18}$ & $2.50773\times 10^{-18} $ & $1.99217\times 10^{-18}$ & $9.61791\times 10^{-19}$ & $6.77894\times 10^{-19}$  & $4.15999\times 10^{-19}$ & $2.50497\times 10^{-19}$\\

$n=300$ & $4.32385\times 10^{-18}$ & $2.52126\times 10^{-18} $ & $2.00071\times 10^{-18}$ & $9.63800\times 10^{-19}$ & $6.78908\times 10^{-19}$  & $4.16397\times 10^{-19}$ & $2.50652\times 10^{-19}$\\
\hline

$\alpha^{\beta}_A$ & $6.58123\times 10^{-19}$ & $6.81776\times 10^{-19}$ & $6.67881\times 10^{-19}$ & $5.71538\times 10^{-19}$ & $5.12927\times 10^{-19}$ & $4.31307\times 10^{-19}$ & $3.54102\times 10^{-19}$\\

$n=100$ & $9.76126\times 10^{-19}$ & $9.19324\times 10^{-19}$ & $8.73969\times 10^{-19}$ & $6.98875\times 10^{-19}$ & $6.13252\times 10^{-19}$ & $5.03289\times 10^{-19}$ & $4.05431\times 10^{-19}$\\

$n=150$ & $1.38491\times 10^{-18}$ & $1.20865\times 10^{-18}$ & $1.12119\times 10^{-18}$ & $8.47234\times 10^{-19}$ & $7.29262\times 10^{-19}$ & $5.85979\times 10^{-19}$ & $4.64169\times 10^{-19}$\\

$n=300$ & $2.15163\times 10^{-18}$ & $1.72895\times 10^{-18}$ & $1.56064\times 10^{-18}$ & $1.10529\times 10^{-18}$ & $9.29960\times 10^{-19}$ & $7.28372\times 10^{-19}$ & $5.65045\times 10^{-19}$\\
\hline

$\tilde{\alpha}^\beta_A$ & $6.58123\times 10^{-19}$ & $6.81776\times 10^{-19}$ & $6.67881\times 10^{-19}$ & $5.71538\times 10^{-19}$ & $5.12927\times 10^{-19}$ & $4.31307\times 10^{-19}$ & $3.54081\times 10^{-19}$\\

$\alpha^\beta_A-\tilde{\alpha}^\beta_A$ & $-6.5002\times 10^{-29}$ & $-9.5531\times 10^{-29}$ & $-1.1046\times 10^{-28}$ & $-1.5258\times 10^{-28}$ & $-1.5704\times10^{-28}$ & $1.1085\times 10^{-26}$ & $2.0858\times10^{-23}$\\

$n=100$ & $9.76126\times 10^{-19}$ & $9.19324\times 10^{-19}$ & $8.73969\times 10^{-19}$ & $6.98875\times 10^{-19}$ & $6.13252\times 10^{-19}$ & $5.03289\times 10^{-19}$ & $4.05411\times 10^{-19}$\\

$\alpha^\beta_A-\tilde{\alpha}^\beta_A$ & $ -6.4435\times 10^{-29}$ & $-9.5275\times 10^{-29}$ & $-1.1028\times 10^{-28}$ & $-1.5252\times 10^{-28}$ & $-1.5701\times 10^{-28}$ & $1.1085\times 10^{-26}$ & $2.0858\times 10^{-23}$\\
\hline

$\beta_A$ & $4.82093\times 10^{-18}$ & $3.14486\times 10^{-18}$ & $2.63118\times 10^{-18}$ & $1.52608\times 10^{-18}$ & $1.18708\times 10^{-18}$ & $8.4580\times 10^{-19}$ & $6.040003\times 10^{-19}$\\

$n=100$ & $5.21066\times 10^{-18}$ & $3.40926\times 10^{-18}$ & $2.85475\times 10^{-18}$ & $1.65787\times 10^{-18}$ & $1.28972\times 10^{-18}$ & $9.18719\times 10^{-19}$ & $6.55704\times 10^{-19}$\\

$n=150$ & $5.66915\times 10^{-18}$ & $3.71638\times 10^{-18}$ & $3.11337\times 10^{-18}$ & $1.80902\times 10^{-18}$ & $1.40716\times 10^{-18}$ & $1.00198\times 10^{-18}$ & $7.14666\times 10^{-19}$\\

$n=300$ & $6.47549\times 10^{-18}$ & $4.25021\times 10^{-18}$ & $3.56135\times 10^{-18}$ & $2.06909\times 10^{-18}$ & $1.60887\times 10^{-18}$ & $1.14477\times 10^{-18}$ & $8.15697\times 10^{-19}$\\
\hline

$\alpha_B$ & $3.21327\times 10^{-18}$ & $1.84256\times 10^{-18}$ & $1.44482\times 10^{-18}$ & $6.57662\times 10^{-19}$ & $4.46029\times 10^{-19}$ & $2.56302\times 10^{-19}$ & $1.42028\times 10^{-19}$\\

$n=100$ & $3.28514\times 10^{-18}$ & $1.86946\times 10^{-18}$ & $1.46230\times 10^{-18}$ & $6.62119\times 10^{-19}$ & $4.48344\times 10^{-19}$ & $2.57239\times 10^{-19}$ & $1.42403\times 10^{-19}$\\

$n=150$ & $3.33485\times 10^{-18}$ & $1.88725\times 10^{-18}$ & $1.47369\times 10^{-18}$ & $6.64911\times 10^{-19}$ & $4.49774\times 10^{-19}$ & $2.57809\times 10^{-19}$ & $1.42627\times 10^{-19}$\\

$n=300$ & $3.37446\times 10^{-18}$ & $1.90078\times 10^{-18}$ & $1.48223\times 10^{-18}$ & $6.66917\times 10^{-19}$ & $4.50792\times 10^{-19}$ & $2.58204\times 10^{-19}$ & $1.42779\times 10^{-19}$\\
\hline
\hline

$\alpha_B^{S}$ & $3.39012\times 10^{-18}$ & $1.90823\times 10^{-18}$ & $1.48794\times 10^{-18}$ & $6.68541\times 10^{-19}$ & $4.51217\times 10^{-19}$ & $2.57978\times 10^{-19}$ & $1.42812\times 10^{-19}$\\

$n=100$ & $3.1 \%$ & $2.0 \%$ & $1.7 \%$ & $0.9 \%$ & $0.6 \%$ & $0.3\%$ & $0.3\%$\\

$n=150$& $1.6 \%$ & $1.1 \%$ & $0.9 \%$ & $0.5 \%$ & $0.3 \%$ & $0.07\%$ & $0.12\%$\\

$n=300$& $0.46 \%$ & $0.39 \%$ & $0.38 \%$ & $0.24 \%$ & $0.09 \%$ & $-0.09\%$ & $0.02\%$\\

 \hline
  \hline
\end{tabular}
\end{table}
\end{center}
\end{widetext}
 %, \textbf{ wherefrom it follows that the account for the finite lifetimes of atomic levels does not affect on determination of the cosmological parameters.}
The two cases (A and B) of astrophysical investigations are noted by the corresponding indexes. The case B can be easily obtained by the subtraction of corresponding values of $\alpha_{1s}$ from $\alpha_A$. The values of $\alpha_B$ are compared in the last rows of Table~\ref{tab:2} with the results calculated via the extrapolation formula, see \cite{Seager_2000} and references therein:
\begin{eqnarray}
\label{40}
\alpha^S_B=10^{-19}\frac{a\, t^b}{1+c\, t^d}{\rm m^3 s^{-1}},
\end{eqnarray}
where $a = 4.309$, $b = -0.6166$, $c = 0.6703$ and $d = 0.5300$. 

Relative difference of our calculations and results obtained with Eq. (\ref{40}) for $\alpha_B$ is about $5.2\%$ at the temperature $300$ K and decreases to $0.55\%$ at the temperature $20000$ K for the set of $nl$ states with $n_{\rm max}=70$. The deviation occurs due to the two curcumstances: {\rm i}) at low temperatures the larger number of states $nl$ should be taken into account in view of bad convergence of sum over $nl$ set; {\rm ii}) the extrapolation formula does not work well at low temperatures. The discrepancy between $\alpha_B$ and $\alpha^S_B$ decreases with the account for the larger set of quantum states, the relative difference in percents is given in the last rows of Table~\ref{tab:2} and does not exceed the $0.5\%$ at $n_{\rm max}=300$. Applying the same fit as Eq. (\ref{40}) to our data for the set of quantum states with $n_{\rm max}=300$, i.e. taking into account 45000 atomic levels, one can find $a = 4.2707$, $b = -0.6172$, $c = 0.6554$ and $d = 0.5307$. It is expected that such fit modification can lead to the correction of the order of $0.25\%$ for the ionization fraction of primordial plasma. 

An accuracy of our calculations can be checked via the detailed balance relation. According to this the sum of spontaneous and stimulated recombination coefficient should be equal to the photoinization coefficient: $\alpha_A+\alpha_A^\beta=\beta_A$. In our calculations the largest relative difference corresponds to the temperature $300$ K and is about $5.4\times 10^{-13}$, whereas for the temperature $20000$ K it is about $4.5\times 10^{-17}$. Such relative deviation allows us to analyse the effect of finite lifetimes on the photorecombination process. For this purpose we compare $\alpha_A^\beta$ and $\tilde{\alpha}_A^\beta$ defined by Eqs.  (\ref{31}), (\ref{34}) and (\ref{38}), respectively. As it follows from Table~\ref{tab:2} the magnitude of $\tilde{\alpha}^\beta_A$ is almost the same as $\alpha_A^\beta$. The more or less significant difference arises at temperatures $10000$ K and $20000$ K, where the deviation in seventh and fourth digit is found, respectively. We demonstrate the difference but not the relative to show that this magnitude is 'stable' for any set of atomic states. Thus, we do not calculate $\tilde{\alpha}^\beta_A$ for the $n_{\rm max}$ larger than $100$. Since the cross-section for the stimulated recombination process increases with the account for the larger set of $nl$ states the relative difference $(\alpha_A^\beta-\tilde{\alpha}_A^\beta)/\alpha_A^\beta$ becomes less. Nonetheless, we can evaluate the relative difference $(\alpha_A^\beta-\tilde{\alpha}_A^\beta)/\alpha_A$ which is about $0.0083\%$ at temperature $20000$ K and becomes negligible at lower temperatures. Thus, we can conclude that the effect of finite lifetimes of atomic levels leads to violation of the detailed balance relation on the level of $0.01\%$ for high temperatures and can be important at low temperatures for the investigations where only the partial unbound-bound transitions play the role, see Table~\ref{tab:1}.

\section{Conclusions}
In this paper the new closed expression for evaluating of photorecombination and photoionization cross-section is suggested within the rigorous Quantum Electrodynamics approach. This expression allows for the accurate (unphenomenologial) introducing of the finite lifetimes of atomic levels. This leads to appearnce of level widths in the expression for the stimulated photorecombination cross-section, see Eq. (\ref{34}). Results of Table~\ref{tab:1} show that this effect can be important in investigations where the partial photorecombination processes have a meaning.

Moreover, the total recombination and ionization coefficients as well as stimulated photorecombination coefficient are evaluated. It is shown that values of total recombination and ionization coefficients are close to the results obtain with the extrapolation formula Eq. (\ref{40}). However, the difference of these two magnitudes increases with the descreasing of temperature, see Table~\ref{tab:2}. This happens in view of slow convergence of total photorecombination coefficient for low temperatures and series of approximations used in derivation of Eq. (\ref{40}). Therefore, we have used our data to find the new extrapolation coefficients: $a = 4.2707$, $b = -0.6172$, $c = 0.6554$ and $d = 0.5307$. The rough estimations show that such modificantion of $\alpha_B$ coefficient can lead to $0.25\%$ contribution in ionization fraction of primordial plasma. 

Finally, the results listed in Table~\ref{tab:2} show that the effect of finite lifetimes of atomic levels leads to violation of detailed balance principle: $\alpha_A+\alpha_A^\beta=\beta_A$. In our calculations this relation is fulfilled on the level of $10^{-17}$ at high temperatures and $10^{-13}$ at the temperature $300$ K within the Quantum Mechanical approach. The account for atomic level widths (QED effect) results in appearance of Lorentz profile instead of $\delta$-function, see Eq. (\ref{35}). In turn, such modification stretches the distribution function under the integral Eq. (\ref{34}). The comparison of QM and QED results are given in Table~\ref{tab:2} as the difference of $\alpha_A^\beta-\tilde{\alpha}_A^\beta$. In particular, from Table~\ref{tab:2} follows that values of the stimulated photorecombination coefficient are larger within QED approach at low temperatures and less than QM results at high temperatures, $\alpha_A^\beta\approx\tilde{\alpha}_A^\beta$ at the temperature about $7500$ K. The magnitude $\alpha_A^\beta-\tilde{\alpha}_A^\beta$ is the same for the different sets of quantum states at fixed temperature. The relative difference with respect to $\alpha_A$ is about $0.0083\%$ at temperature $20000$ K what demonstrates the violation of detailed balance principle.
\section{Acknowledgments}

This work was supported by Russian Science Foundation (grant 17-12-01035).

\bibliography{mybibfile}

\end{document}